\documentclass[sigconf]{acmart}
\usepackage{booktabs} 
\usepackage{subcaption}
\usepackage{listings}
\usepackage{threeparttable}
\usepackage{multirow}
\usepackage{mwe,tikz}
\usepackage{siunitx}
\usepackage{xparse}
\usepackage{enumitem}

\newsavebox{\fminipagebox}
\NewDocumentEnvironment{fminipage}{m O{\fboxsep}}
 {\par\kern#2\noindent\begin{lrbox}{\fminipagebox}
  \begin{minipage}{#1}\ignorespaces}
 {\end{minipage}\end{lrbox}%
  \makebox[#1]{%
    \kern\dimexpr-\fboxsep-\fboxrule\relax
    \fbox{\usebox{\fminipagebox}}%
    \kern\dimexpr-\fboxsep-\fboxrule\relax
  }\par\kern#2
 }

 \definecolor{deepred}{rgb}{0.6,0,0}
 \definecolor{deepgreen}{rgb}{0,0.5,0}

\lstdefinestyle{PythonStyle}
{
        language=Python,
        basicstyle=\ttfamily,
        upquote=true,
        numbers = none,
        numberstyle=\footnotesize,
        escapechar=`,
        float,
        moredelim=[il][]{--latexlabel},
        otherkeywords={self},             
        commentstyle=\color{blue},
        keywordstyle=\bfseries\color{black},
        emph={MyClass,__init__,Config,HighThroughputExecutor,SlurmProvider,LocalChannel}, 
        emphstyle=\bfseries\color{deepred},    
        showstringspaces=false            %
}

\lstdefinestyle{PythonStyleInLine}
{
        language=Python,
        basicstyle=\ttfamily,
        upquote=true,
        numbers = none,
        numberstyle=\footnotesize,
        escapechar=`,
        moredelim=[il][]{--latexlabel},
        otherkeywords={self},             
        commentstyle=\color{blue},
        keywordstyle=\bfseries\color{black},
        emph={MyClass,__init__,@python_app,@bash_app},          
        emphstyle=\bfseries\color{deepred},    
        showstringspaces=false            %
}

\newif\iffinal

\finaltrue

\iffinal
  \newcommand{\ian}[1]{}
  \newcommand{\ryan}[1]{}
  \newcommand{\kyle}[1]{}
  \newcommand{\zhuozhao}[1]{}
  \newcommand{\katznote}[1]{}
  \newcommand{\benc}[1]{}
  \newcommand{\yadu}[1]{}
  \newcommand{\rohan}[1]{}
\else
  \newcommand{\ian}[1]{{\textcolor{red}{ Ian: #1 }}}
  \newcommand{\ryan}[1]{{\textcolor{magenta}{ Ryan: #1 }}}
  \newcommand{\kyle}[1]{{\textcolor{purple}{ Kyle: #1 }}}
  \newcommand{\zhuozhao}[1]{{\textcolor{blue}{ Zhuozhao: #1 }}}
  \definecolor{darkgreen}{rgb}{0,0.5,0}
  \newcommand{\katznote}[1]{{\textcolor{darkgreen}{ Dan: #1 }}}
  \definecolor{darkpurple}{rgb}{0.5,0,0.5}
  \newcommand{\benc}[1]{{\textcolor{darkpurple}{ Ben: #1 }}}
  \newcommand{\yadu}[1]{{\textcolor{orange}{ Yadu: #1 }}}
  \definecolor{pink}{rgb}{1.0,0,0.5}
  \newcommand{\rohan}[1]{{\textcolor{pink}{ Rohan: #1 }}}
\fi

\renewcommand\footnotetextcopyrightpermission[1]{}

\usepackage{fancyhdr}
\fancypagestyle{firststyle}
{
	\fancyhf{}
	\fancyhead[C]{\large This is a preprint of an HPDC-28 paper, which will be published as \url{https://doi.org/10.1145/3307681.3325400}.}
	
}

\begin{document}
\title{Parsl: Pervasive Parallel Programming in Python}

\author{Yadu Babuji}
\affiliation{%
  \institution{University of Chicago}
}
\email{yadunand@uchicago.edu}

\author{Anna Woodard}
\affiliation{%
  \institution{University of Chicago}
}
\email{annawoodard@uchicago.edu}

\author{Zhuozhao Li}
\affiliation{%
  \institution{University of Chicago}
}
\email{zhuozhao@uchicago.edu}

\author{Daniel S. Katz}
\affiliation{%
  \institution{University of Illinois at Urbana-Champaign}
}
\email{d.katz@ieee.org}

\author{Ben Clifford}
\affiliation{%
  \institution{University of Chicago}
}
\email{bzc@uchicago.edu}

\author{Rohan Kumar}
\affiliation{%
  \institution{University of Chicago}
}
\email{rohankumar@uchicago.edu}

\author{Lukasz Lacinski}
\affiliation{%
  \institution{University of Chicago}
}
\email{lukasz@uchicago.edu}

\author{Ryan Chard}
\affiliation{%
  \institution{Argonne National Laboratory}
}
\email{rchard@anl.gov}

\author{Justin M. Wozniak}
\affiliation{%
  \institution{Argonne National Laboratory}
}
\email{woz@anl.gov}

\author{Ian Foster}
\affiliation{%
  \institution{Argonne \& U.Chicago}
}
\email{foster@anl.gov}

\author{Michael Wilde}
\affiliation{%
  \institution{ParallelWorks}
}
\email{wilde@parallelworks.com}

\author{Kyle Chard}
\affiliation{%
  \institution{University of Chicago}
}
\email{chard@uchicago.edu}

\renewcommand{\shortauthors}{Y. Babuji et al.}

\begin{abstract}
High-level programming languages such as Python are increasingly used to provide 
intuitive interfaces to libraries written in lower-level languages
and for assembling applications from various
components. This migration towards orchestration rather than implementation, 
coupled with the growing need for parallel computing (e.g., due to big 
data and the end of Moore's law), necessitates rethinking how parallelism is expressed in programs.
Here, we present Parsl, a parallel scripting library that augments 
Python with simple, scalable, and flexible constructs for encoding 
parallelism. These constructs allow Parsl 
to construct a dynamic dependency graph of 
components that it can then execute efficiently on one or many processors.
Parsl is designed for scalability, with an extensible
set of executors tailored
to different use cases, such as low-latency, high-throughput, or
extreme-scale execution. We show, via experiments on the Blue Waters
supercomputer, that Parsl executors can allow Python scripts to execute
components with as little as 5 ms of overhead, scale 
to more than \num{250000} workers across more than \num{8000} nodes, 
and process upward of \num{1200} tasks 
per second. Other Parsl features simplify
the construction and execution of composite programs by supporting
elastic provisioning and scaling of infrastructure, fault-tolerant execution, and 
integrated wide-area data management. 
We show that these capabilities satisfy the needs 
of many-task, interactive, online, and machine learning applications in 
fields such as biology, cosmology, and materials science.
\end{abstract}

%
%
%
%


\maketitle

\thispagestyle{firststyle}

\section{Introduction}

The past decade has seen a major transformation in the nature of programming. 
Software is increasingly constructed by using a high-level language to integrate 
components from many sources. In other words, much software is not so much written as assembled.
Additionally, as data sizes increase
and sequential processing power plateaus
there is a growing need to make use of
parallel hardware such as specialized accelerators and distributed computing systems.
As a contribution to a merging of these two trends, 
we present here methods that allow for the natural expression of parallelism 
within a popular high-level language, Python,
in such a way that programs can express opportunities for parallelism that can then 
be realized, at execution time, using different execution models on different parallel platforms.

Specifically, we present Parsl~\cite{parsl}, a parallel scripting library that defines and implements
Python decorators that developers can use to express parallelism with Python programs.
We show how programmers can thus easily
create programs composed of both Python functions and components 
written in other languages, with opportunities for parallel execution expressed in a
way that allows for efficient implementation on a variety of architectures. 
Parsl thus implements a form of compositionality, in which a
program is constructed by composing component programs in parallel. 
This approach to parallelism is in contrast to prior efforts that 
rely on domain-specific languages (DSL)~\cite{foster1992productive,swift,di2017nextflow}, configuration-based models~\cite{pegasus,galaxy,cwl}, Python-based graph descriptions~\cite{jain2015fireworks, luigi}, 
and compiled language extensions~\cite{mani92cc++} to support such composition. 
We choose to extend Python for 
several reasons: it is widely used for programming;
as an interpreted language it can be easily extended (without requiring compiler modifications);
and while not considered a functional programming language, 
it lends itself well to a functional style that fits our model of parallelism.

Parsl enables a simple functional programming model at the task level while 
still retaining procedural Python code for other aspects of the program. 
That is, it allows developers to declare the logic of a program,
without explicitly describing how components are executed, by chaining together
functions with defined input and output objects. 
Developers simply annotate functions in a Python script as \emph{Apps},
indicating that these functions (either pure Python or components in 
other languages) can be executed concurrently, if permitted by data dependencies.
Parsl composes a dynamic dependency graph and maps the task
graph to arbitrary execution resources, exploiting parallelism where possible by 
tracking the creation of data (objects or files) and 
determining when App dependencies are met. 
The explicit input/output specification for Apps, immutable input and output
object passing, and dependency-based
execution model together enable users to construct safe and deterministic
parallel programs. 
Parsl is inherently flexible and scalable. Its modular execution architecture
supports a variety of execution models---from pilot jobs to extreme
scale distributed execution---and a variety of execution providers, from laptops
to supercomputers. By separating code from configuration,
Parsl allows parallelism to be realized in different ways on different
resources, without changing program code. 
Together, these features enable simple, safe, scalable, and flexible parallelism.

The fact that Parsl extends Python offers many advantages over alternative approaches.
For example, Parsl can be easily configured
on a variety of computers (including in virtual Python environments),
as Python is part of standard distributions and is already deployed in many environments.
Parsl allows users to express parallelism in familiar
Python code, without needing to learn new syntax.
Developers familiar with Python can, in a matter of minutes, 
learn the additional constructs offered by Parsl. 
Furthermore, Parsl programmers are not constrained to using only Parsl constructs:
they can access the complete power of Python to 
implement sequential functionality (e.g., argument parsing, logging, visualization
libraries), leverage the vibrant and diverse  scientific Python (SciPy)
ecosystem (e.g., Pandas dataframes, Scikit-learn, 
Jupyter notebooks, and Python interfaces to other tools like Tensorflow), 
and exploit powerful language features  
such as conditional and loop constructs, complex datatypes, generators, 
list comprehensions, and other object-oriented functionality. This
flexibility is important because there is usually much more to a program 
than parallelism. Parsl embraces this idea, making it easy to augment
Python programs with Parsl-based parallel components, rather than
relying on complex nesting
or specifications defined in terms of markup languages.

In this paper we describe our motivation for developing Parsl, 
the design decisions that influenced its design, and its implementation architecture. 
We evaluate the scalability of Parsl on a campus cluster and a supercomputer,
showing that its high-throughput executor can scale to more than \num{65000} concurrent workers
with throughput greater than \num{1000} tasks/second, 
that its extreme-scale executor can scale to more than \num{250000} concurrent
workers over \num{8000} nodes, and that its low-latency executor can execute components within 5ms, far
exceeding what other Python-based systems can achieve. 

This paper is structured as follows. In \S\ref{sec:motivation} we outline
five use cases that motivated the development of Parsl. 
In \S\ref{sec:design} and \S\ref{sec:architecture} we outline the design
and architecture of Parsl, respectively. In \S\ref{sec:evaluation} 
we evaluate the scalability and performance of Parsl.
We present related work in \S\ref{sec:related}.
Finally, in
\S\ref{sec:summary} we summarize our contributions.

\section{Motivation}
\label{sec:motivation}

To motivate our approach we first highlight five scientific use cases
that exhibit a variety of requirements, including the management of progress, concurrency, and data.  We then describe
the advantages of implementing Parsl as an extension to Python. 

\subsection{Use Cases}
\label{sec:usecases}
Parsl has been designed to support a broad range of science and engineering use cases, 
from traditional many-task computing to new computational modes such as
interactive computing (e.g., in Jupyter notebooks),
machine learning (training and inference), and online computing. 
Here we outline several of the scientific use cases that collectively influenced 
Parsl's architecture, and
we summarize the requirements for each use
case in Table~\ref{tab:use_cases}.

	\begin{table*}[t]
\small
    \centering
\caption{Requirements from five different scientific use cases. 
HTC=High Throughput Computing; FaaS=Function as a Service.}
\vspace{-3mm}
\begin{tabular}{ | l | c| c | c|c|c|}
\hline
 & \textbf{Sequence analysis}     & \textbf{ML inference} & \textbf{Materials science} & \textbf{Neuroscience} & \textbf{Cosmology}  \\ \hline
\textbf{Pattern}       & dataflow     & bag-of-tasks & dataflow & sequential & dataflow \\ \hline
\textbf{Paradigm}       & HTC         & FaaS     & Interactive & Batch     & HTC \\ \hline
\textbf{O(Nodes)}       & hundreds   & tens & tens  & tens   & thousands   \\ \hline
\textbf{O(Tasks)}       & thousands   & thousands         & hundreds & hundreds           & millions   \\ \hline
\textbf{O(Duration)}    & days  & seconds & minutes  & hours  & day  \\ \hline
\textbf{O(Data)}        & TB          & KB  & MB   & GB        & TB \\ \hline
\textbf{Latency Sensitive}& no        & yes      & yes       & no        & no \\ \hline
\end{tabular}
	\label{tab:use_cases}
    \vspace{-2mm}
	\end{table*}

A common many-task application is DNA \textbf{sequence analysis},
which is computationally-intensive,
data-intensive, and requires multiple processing steps using various processing tools (e.g., 
alignment, quality control, variant calling, etc.) One specific DNA sequencing analysis application
is SwiftSeq~\cite{pitt2017deciphering}, which combines many
processing tools and performs highly parallel execution on clouds, clusters, and supercomputers. 
SwiftSeq allows researchers to specify analysis requirements (e.g., 
analysis type, tools to be used, tool parameters) and then dynamically generates 
a many-task workflow for processing. 
SwiftSeq is implemented in Python, and requires a simple way of expressing
parallelism for its processing tools and sequencing data. It is intended
for analysis of thousands of genomes, each several GB in size, 
and with processing tools that run for minutes to hours. 
These long-running tools require fault tolerance.
Efficient use of infrastructure requires that individual tasks be partitioned, 
where possible, and placed across multiple nodes and even sites.

An example of a bag-of-tasks application is \textbf{ML inference}. Increasingly, 
science is performed through multi-tenant, service-oriented platforms, such as scientific portals and gateways.
These services provide on-demand access to the data, tools, and infrastructure required for thousands of researchers
to concurrently perform analyses. 
The Data and Learning Hub for science (DLHub)~\cite{chard19dlhub} is one such service designed to publish and serve ML models for parallel inference by a community of researchers.
DLHub requires methods to manage many short-duration inference requests
using a bag-of-tasks execution model. Unlike other bag-of-tasks workloads, 
DLHub is used in a variety of real-time workloads that require low-latency
responses for example to detect errors. Finally, task isolation, via 
containers, is desirable to accommodate the vanguard requirements of diverse ML models. 
 
An example of interactive computing in \textbf{materials science} is modeling stopping power
in Jupyter notebooks.
Because traditional time-dependent density functional theory (TD-DFT) computations are expensive, 
researchers are developing machine learning 
methods to augment TD-DFT simulations. For example, by using
existing DFT data stored in the Materials Data Facility (MDF)~\cite{blaiszik2016materials} to create 
surrogate models that predict stopping power in different directions.  
During the model development phase, researchers use Jupyter notebooks to 
iteratively develop and evaluate their models. They therefore
require methods that make it easy to parallelize these processes, 
low-latency responses when exploring modeling approaches in an interactive manner, 
and scalability to exploit HPC resources for final model execution.

Experimental research processes often integrate computational analyses for 
error checking, quality control, and  experiment steering. 
For example, \textbf{neuroscience} researchers use x-ray microtomography at the Advanced Photon Source
to characterize the neuroanatomical structure of brain volumes 
to study brain aging and disease~\cite{chard2018high}. 
They combine analysis processes, in near-real time, to reconstruct 
3-dimensional images for error detection and sample orientation during the experiment.
Such analyses are implemented by first identifying the center
of imaged samples from amongst hundreds of 2-dimensional slices, 
applying a machine learning model to identify the highest quality slices, 
and then using tomographic reconstruction to create a 3D model from 
the slices. 
To reliably perform such reconstructions in a timely manner, researchers must be able to leverage
the largest computing resources available to them, such as those at
Argonne's Leadership Computing Facility. 

Analyses can also have unpredictable performance.
An example is a \textbf{cosmology} simulation 
that aims to produce simulated images from the Large Synoptic Survey Telescope (LSST). 
To create these simulated images, researchers first construct more than \num{10000}
instance catalogs of cosmological objects using observation parameters (e.g., time, 
altitude, temperature, telescope configuration) and astrophysical inputs
(e.g., stars, galaxies).
They then use these catalogs to simulate images acquired from each of the LSST's 189 sensors.
To simplify development and portability, the simulation code is written in Python
and packaged in Singularity containers.  
Given that the simulation can occupy the full capacity of a supercomputer 
for weeks, the parallel execution system
must be capable of maintaining high utilization at scale. For example,
as execution time is dependent on the number of objects included in a sensor/catalog, there
is potential for significant imbalance throughout the simulation, thus the simulation must group
(and rebalance) tasks into appropriate sized bundles for a given processing node (e.g., 64 tasks for a 64-core processor).

\subsection{Why build on Python?}
Python, first released in 1991, has become the \textbf{lingua franca} in many domains.
Many developers, scientists, and analysts use Python extensively as 
it is straightforward to learn, well documented, and reliable.
Python is a powerful programming language with 
sophisticated features that need neither be reimplemented by Parsl nor
relearned by programmers: for example, if statements, loops, generators, objects, and list comprehensions.
Python also has a rich and vibrant ecosystem with many useful
libraries.

We briefly outline here some of the advantages of implementing Parsl
as a library for Python. 

\textbf{Parsl semantics differ from those of Python only where necessary.}
Everything else stays the same and need not be designed,
implemented, or learned. 
Thus, Parsl is familiar to Python developers and can be easily learned by others.
Although Parsl focuses on task-oriented parallel computation, a
program needs to do other things---often relatively
trivial booking-keeping work that should rightly be implemented easily.
As a tradeoff, reasoning about the program as a
whole can be harder (affecting the correctness of checkpointing and retries,
for example).

\textbf{A single language for the implementation of Parsl and for writing programs.} 
Parsl blurs the line between library and programs written with 
the library. For example, the LSST simulation above can use
a small, straightforward piece of Python code to 
rate limit and rewrite the program's work queue and thus
influence Parsl's scheduling in a non-trivial,
program-specific manner.
Such behavior is not expressible as part of the task
dependency graph; but being application specific, neither would it be
appropriate for it to be part of the core Parsl library.

\textbf{Programming language features are beneficial for programming in the large.} 
Like any other Python code, Parsl can be used to create programs,
and standard libraries of Parsl Apps for particular domains can be developed and
shared. These libraries can then be easily be composed with other code
to create new programs.
The Parsl runtime can execute code from
many such libraries, keeping all the benefits of Parsl task execution
for the program as a whole. Parsl apps look like Python
function calls; asynchronously computed result values look
like Python values and can be passed around and stored as such.
Artificial barriers caused by one component of a program ending and
another beginning can be avoided: as the whole program
can be written in one language, the runtime can manage the execution
of all tasks, no matter the component from which those tasks originates,
and can run tasks from different components in parallel on the same resource.

\textbf{Programming language features are useful for programming in the
small.}
As features develop, domain-specific languages designed to manage
progress and concurrency may struggle to solve simple problems
that are easily managed in sequential scripting languages.
For example, a DAG-based language
may need to choose a later action based on an earlier result.
In a generic imperative programming
language, a simple \texttt{if} statement suffices; a 
DAG-based approach to the same problem may require generating workflows inside
other workflows, or embedding a richer language deep inside the
outer DAG description.
Similarly, a loop over a dynamically generated dataset can be accomplished
with a simple \texttt{for}
statement in Python; in a DAG-based language, it might require a 
separate workflow generation stage that cannot be scheduled as part of the main workflow.
It is access to these Python language constructs that contributes
to Parsl's generality and learnability.

\section{Design}\label{sec:design}

Parsl is designed to enable intuitive parallel and compositional programming in Python
that address broad scientific use cases, such as those
outlined in the previous section. 
Specifically, we focus on five core design challenges: 
enabling the intuitive description of parallelism in Python; 
decomposing dataflow dependencies into a dynamic task graph for efficient execution;
abstracting task execution mechanisms and environments to enable portability;
supporting execution of heterogeneous workloads that range from short-running to long-running tasks, from few to millions of tasks, and from small-scale to large-scale resources; and
enabling reliable parallel program execution.

\subsection{Programming with Parsl}
Parsl is designed for Python. When facing design decisions, 
we have chosen to remain as close to standard Python as possible, 
ensuring that Parsl is easy to learn and that Parsl scripts remain Pythonic.

At the core of Parsl are two constructs that introduce asynchronicity
into Python: The \texttt{App} decorator and \texttt{future} object
that can be used together to compose programs.
These constructs allow Python functions to be executed asynchronously, in parallel, 
and potentially in a different execution location. 

\subsubsection{Apps}

Parsl uses decorators to intercept and modify the behavior of Python functions.
In Parsl two kinds of decorators are supported: the \texttt{@python\_app} decorator
for pure Python functions and \texttt{@bash\_app} decorator for shell commands.

When either a Python App or Bash App is invoked, an asynchronous task is registered with Parsl and
a \texttt{future} object is returned immediately, in lieu of the results of the computation. Eventually
Parsl executes the task and results are made available through the \texttt{future}.
To guarantee safety in a concurrent setting, Parsl Apps must be pure functions, acting only on their
input arguments. The following is a minimal example of a Python App.
\begin{lstlisting}[style=PythonStyleInLine, basicstyle = \small\ttfamily]
@python_app
def hello1(name):
    return "Hello {}".format(name)
\end{lstlisting}

As for Bash Apps, the Python code that forms the body of the function
should return a fragment of Bash shell code. That shell code will be executed 
in a sandbox environment.
Input and output handling behaves as with Python apps, although the return value 
from Bash Apps are UNIX return codes
that indicate only whether the code succeeded.
The Bash App also can be further configured using special keywords that
allow for redirection of STDOUT/STDERR streams to files.
The following is an example.
\begin{lstlisting}[style=PythonStyleInLine, basicstyle = \small\ttfamily]
@bash_app
def hello2(name, stdout=None, stderr=None):
     return "echo 'Hello {}'".format(name)
\end{lstlisting}

\noindent
Apps of either type can be invoked with standard Python syntax:
\begin{lstlisting}[style=PythonStyleInLine, basicstyle = \small\ttfamily]
f1 = hello1("World")
f2 = hello2("World")
\end{lstlisting}

\noindent Thus, Parsl App functions invoked this way automatically follow Parsl semantics, without Parsl syntax at the call site.  This is achieved through Parsl's use of futures and a parallel runtime.

\subsubsection{Futures}

We saw earlier that invoking Parsl Apps return \texttt{futures}. 
A \texttt{future} is an object that can be used to access
the results of an asynchronous computation. 
Parsl futures implement a synchronous blocking method \texttt{future.result()} that
will wait until the computation has yielded results or raised an exception. 
A non-blocking method, \texttt{future.done()},
immediately returns a boolean that indicates the status of the execution.

Since a \texttt{future} is created and updated only by a specific App invocation, it acts as a single-update
variable, as commonly used in task-parallel systems~\cite{mani92cc++,foster1990strand,foster1992productive,swift}. 
Futures are the only synchronization primitive offered by Parsl.

\subsection{Input and output data}
Any input data required by the App must be passed as input parameters to
the invocation. 
While this represents a familiar way for Python users to pass objects, 
we note that other ways of passing information to Python
function invocations, such as global variables, cannot be
used: a price paid to allow for easy movement of tasks among execution
resources.
Similarly, any output from the App invocation needs to be explicitly
passed back---as a return value or as an output file.

The following three Python types can be passed as inputs or outputs, with caveats:

{\bf Python objects}. Any Python object that can be serialized (e.g., using Python's pickle~\cite{pickle}
or dill~\cite{mckerns2012building} libraries), can be passed as an input parameter. Most Python objects
that represent ``data'' (rather than, for example, file
descriptors or threads) can be serialized.
Objects passed as inputs should be treated as immutable: their contents
should not be modified on either the submitting or
executing side, as non-deterministic behavior can result otherwise.
Parsl does not attempt to provide any richer distributed object model.

{\bf Files} are declared using a Parsl \texttt{File} object, which can
represent a remote file using various protocols. Parsl stages in file inputs 
and translates paths transparently so
that they are available in the runtime environment of the executing program.
Parsl provides convenience keyword arguments \texttt{inputs} and \texttt{outputs}
in App functions that allow developers to dynamically specify collections
of input and output files.
Code should not modify input files, and should not make assumptions about 
visibility of modifications between App invocations: see \S\ref{sec:data}.

{\bf Futures} will eventually contain some value, which may be of any of the above types.  

Parsl is not unnecessarily strict about enforcing these rules: for example, 
if a shared filesystem is available 
Parsl file staging is not necessary. However, if a programmer
chooses to make that tradeoff, the resulting code may not be able to be executed 
in an arbitrary environment.

\subsection{Compositionality}
Parsl allows applications to be composed by passing \texttt{futures} between Apps.
An arbitrary number of \texttt{futures} can be passed to an App as arguments, implicitly
indicating dependencies on the asynchronous execution of all Apps whose futures were passed as arguments.
Since Apps can be invoked asynchronously, 
an arbitrarily large task graph can be constructed
with minimal compute cost. 
Parsl can then execute these tasks based on the available parallelism within
the task graph.

\subsection{Runtime}

Parsl's runtime is responsible for managing the parallel execution of Parsl-annotated
components in a program on configured resources. 
To do so, Parsl assembles and dynamically updates a task dependency
graph. This graph contains all state for the program and can be 
efficiently introspected and mapped to available execution resources. 
The task graph is represented as a directed acyclic graph (DAG) in which
the nodes represent tasks and the edges represent the input/output data exchanged
between tasks. The advantage of using a dynamic task graph is
twofold: first, the execution of tasks may start as soon as the first task is 
identified, rather than waiting for the entire task graph to be formed before execution; 
and second, it allows for complex logic to be implemented in the program (e.g., loops and conditionals) as well
as for tasks to generate new tasks during execution.

\subsection{Separation of Code and Configuration}
Parsl separates program logic
from execution configuration, with the latter described by a Python object so that
developers can easily introspect permissible options, validate
settings, and retrieve/edit configurations dynamically during execution.
A configuration specifies details of the provider, executors, connection
channel, allocation size, queues, durations, and data management options. Listing~\ref{lst:parsl-configuration}
illustrates a basic configuration for the Stampede2 supercomputer. This configuration
uses the HighThroughputExecutor to submit tasks from a login node (LocalChannel).
It requests an allocation of 128 nodes, from the \texttt{skx-normal} partition, for up to
12 hours.
\begin{lstlisting}[style=PythonStyle, basicstyle = \small\ttfamily, label={lst:parsl-configuration}, caption={Parsl configuration for Stampede2.}]
    config = Config(
        executors=[
            HighThroughputExecutor(
                label="stampede2_htex",
                address=address_by_hostname(),
                provider=SlurmProvider(
                    channel=LocalChannel(),
                    nodes_per_block=128,
                    init_blocks=1,
                    partition="skx-normal",
                    walltime="12:00:00"
                )
            )
        ]
    )
\end{lstlisting}

Many use cases require the ability to mix-and-match execution resources
in various ways. For example, using GPU nodes for GPU-optimized codes
and CPU nodes for others; combining thread-based execution of lightweight
tasks with cluster-based execution of larger tasks; or even executing 
tasks across a number of computing resources wherever allocations may
be available. Parsl supports these use cases by enabling ``multi-site''
execution via specification of more than one executor in the configuration.

\subsection{Scalable Execution}

Parsl is designed to support a wide range of use cases, from
few long tasks to millions of short tasks.
A system designed to address any such use case individually is likely to be 
unsuitable for other use cases. 
We focus instead on providing a generic and 
extensible model for supporting these varying use cases.

Parsl's modular executor interface is designed to support
different mechanisms for managing the execution of tasks. The executor
controls the process by which the task is transported 
to configured resources, executed on that resource, and results
are communicated back to Parsl. In \S\ref{sec:executors} we describe
a set of executors that provide high-throughput, low-latency, and extreme-scale execution.

As execution progresses the resources required can vary. 
For example, the map-reduce pattern, common in many-task
workflows, starts with a large set of map tasks, the results of which are combined
by a smaller number of reduce tasks. To minimize resource wastage, Parsl
is designed to automatically provision and deprovision execution resources.
This elasticity component is controlled by an extensible \emph{strategy}
module within Parsl. The strategy module tracks outstanding tasks and available
capacity on connected executors.  It communicates
with the connected providers to automatically
scale to match real-time requirements. 

\subsection{Fault-Tolerant Execution}
A Parsl program may fail due to failure of one of its Apps or of a node 
used for execution. 
As analysis sizes increase, so too does the likelihood of failure. 
In order for Parsl to be usable, it must expect
failures and respond accordingly. For example, 
when re-executing a branch of failed execution, a user is unlikely
to want to re-execute another branch that completed successfully. 

Parsl provides fault-tolerance at the level of programs
rather than Apps.  That is, it treats tasks as the basic
unit for fault tolerance, enabling checkpointing of execution state
whenever a task completes. This ensures that, if desired, a user
may re-execute a program and any Apps that are called with the 
same arguments need not be re-executed. 
While Parsl provides no App-level fault-tolerance, 
it works with Apps that implement their own fault-tolerance, 
e.g., via checkpointing; such methods are opaque
to Parsl. 
A more integrated model of fault-tolerance is future work.
\section{Architecture and implementation}  \label{sec:architecture}

\begin{figure}[h]
 \centering
 \includegraphics[width=0.95\columnwidth]{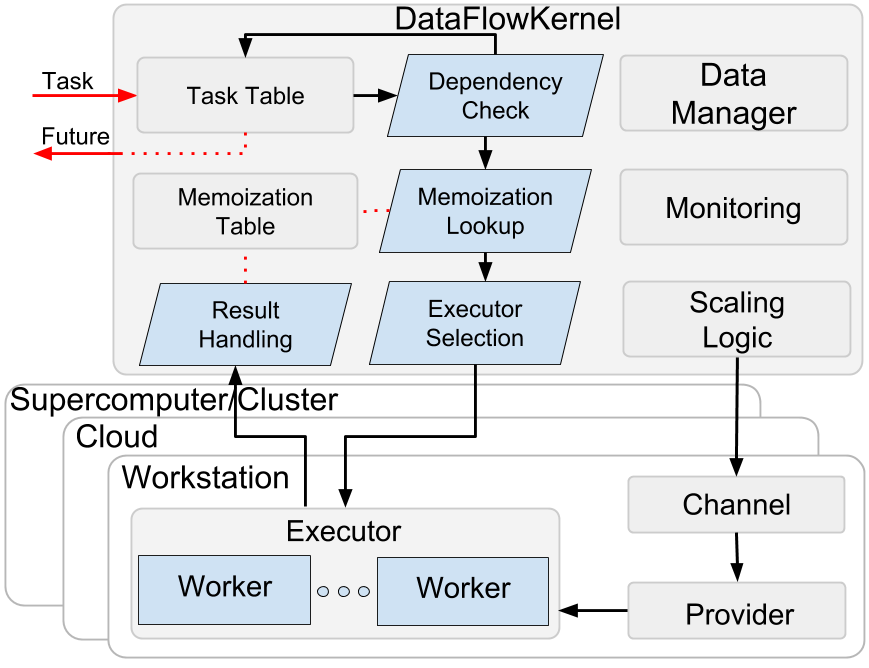}
 
 \vspace{-2ex}
  
 \caption{Parsl architecture. The DataFlowKernel stores the task graph and manages execution
via one or more connected executors, of which three are shown here.}
 \label{fig:parsl-arch}
 \vspace{-2ex}
\end{figure}

Figure~\ref{fig:parsl-arch} shows Parsl's high-level architecture. 
In this section we briefly outline its core components and capabilities. 

\subsection{DataFlowKernel}

The DataFlowKernel (DFK), Parsl's execution management engine, is responsible
for constructing and orchestrating the execution of the task graph. DFK tracks
task information (nodes in the task graph) in a Python data structure. Dependencies
between Apps are implicitly derived from the passing of futures between Apps. 
Edges in the task graph are encoded as asynchronous callbacks on a dependent future,
allowing DFK to be event driven, with each node in the task graph being considered upon resolution
of its edges. 
Launching a task incurs a small fixed cost, as does 
triggering each outgoing edge when a task succeeds.
Thus the execution time complexity of a task graph with $n$ tasks
and $e$ edges is $\mathcal{O}(n + e)$.

Once all of a task's dependencies have resolved successfully, DFK schedules the
task for execution on a configured executor. If multiple executors are available, and
the task contains no execution hints, an executor is picked at random.
The executor responds with its own future that DFK associates with the future that
was created when the App was invoked.

DFK tracks task state and waits for the future to be resolved, or a timeout period to elapse. 
In the case an App fails, as indicated by an exception in the resulting future or timeout, 
Parsl is able to retry the task by resubmitting it to an executor. If retries are disabled, 
or if the number of retries is exceeded, DFK wraps the exception (e.g., App execution 
errors or remote system failure) and associates it with the future. 
When memoization or checkpointing is used, DFK computes a hash of the App's function body and 
performs a lookup in a checkpoint file or memoization table using the function name, body hash, 
and arguments as the key. If the lookup succeeds, the result from the checkpoint file or memoization table is
returned.

\subsection{Providers} 
Clouds, supercomputers, and local PCs offer vastly different modes of access.
To overcome these differences, and present a single
uniform interface, Parsl implements a simple provider abstraction.
This abstraction is key to Parsl's ability to enable scripts to be moved between resources. 
While there have been many prior efforts to implement standard interfaces
for accessing various resources~\cite{saga, drmaa}, 
we chose to develop a lightweight abstraction entirely in Python for Parsl. 
The provider interface is based on three core actions: 
submit a job for execution  (e.g., \texttt{sbatch} for the Slurm resource manager), retrieve the status of an
allocation (e.g., \texttt{squeue}), and cancel a running job (e.g., \texttt{scancel}).
Parsl implements providers for local execution (fork), 
for various cloud platforms using cloud-specific APIs, and for clusters and supercomputers
that use a Local Resource Manager (LRM) to manage access to resources. 
Given the simplicity of the provider interface, it is easy to 
add new providers. Currently, Parsl implements providers for
Slurm, Torque/PBS, HTCondor, Cobalt, GridEngine, AWS, Google Cloud, Jetstream, and Kubernetes. 

Each provider implementation may allow users to specify additional parameters for further
configuration. Parameters are generally mapped to LRM submission script or cloud API options. Examples
of LRM-specific options are partition, wall clock time, scheduler options (e.g., \texttt{\#SBATCH} arguments for Slurm), and worker
initialization commands (e.g., loading a conda environment). 
Cloud parameters include access keys, instance type, and spot bid price.

\subsubsection{Channels}
Parsl scripts may be executed locally (e.g., on a login node or notebook server
at an HPC center) or remotely (e.g., on a laptop, another cluster, or even a laptop). 
To support these different scenarios, Parsl introduces the notion 
of a \emph{Channel} that describes how Parsl should authenticate and connect to the provider.
Parsl includes two primary channels: LocalChannel for execution on a local resource, where
the execution node has direct queue access, and SSHChannel, when executing remotely. 

\subsubsection{Launchers}
Many LRMs offer mechanisms for spawning applications across nodes inside
a single job and for specifying the resources and task placement information
needed to execute that application at launch time. 
Common mechanisms include \texttt{srun} (for Slurm), \texttt{aprun} (for Crays), and \texttt{mpirun} (for MPI). 
The Parsl \emph{Launcher} abstracts
these system-specific launcher systems used to start workers 
across cores and nodes. Users may optionally specify a launcher in the provider
configuration to control how Parsl communicates with the LRM. 
Parsl currently supports the following launchers:
fork, srun, aprun, mpiexec, and GNU parallel.

\subsubsection{Resource management}

One significant challenge when designing a system that makes
use of heterogeneous execution resource types is the need to provide a uniform representation of resources.
Consider that single requests on clouds return individual nodes, clusters and supercomputers provide
batches of nodes, grids provide cores, and workstations provide a single multicore node.
To further complicate matters, some batch systems enforce policies that
restrict the number of nodes permitted in a job (e.g., min, max, or in common groupings). 
From a scheduling perspective, the resources required by each task that is to be scheduled may range from
a fraction of a node through to multiple nodes in the case of MPI applications, 
creating a bin-packing problem.
Parsl defines a resource unit abstraction called a \emph{block} as
the most basic unit of resources to be acquired from a
provider. A block contains one or more nodes and maps to the different
provider abstractions. In a cluster, a block
corresponds to a single allocation request to a scheduler. 
In a cloud, a block corresponds to a single API request for one or more instances. 
Blocks are also used as the basis for elasticity on batch scheduling
systems. Any scaling in/out must occur in units of blocks, as this is the 
most basic unit by which Parsl communicates with the scheduler. This abstraction
can also be used to avoid limitations on the number of nodes that may
be allocated concurrently or on the
number of jobs that may be queued and/or running concurrently.

\vspace{-0.05in}
\subsection{Executors} \label{sec:executors}

\begin{figure*}
\centering
\subcaptionbox{HTEX: High Throughput Executor\label{fig:htex}}{\includegraphics[width=0.28\textwidth]{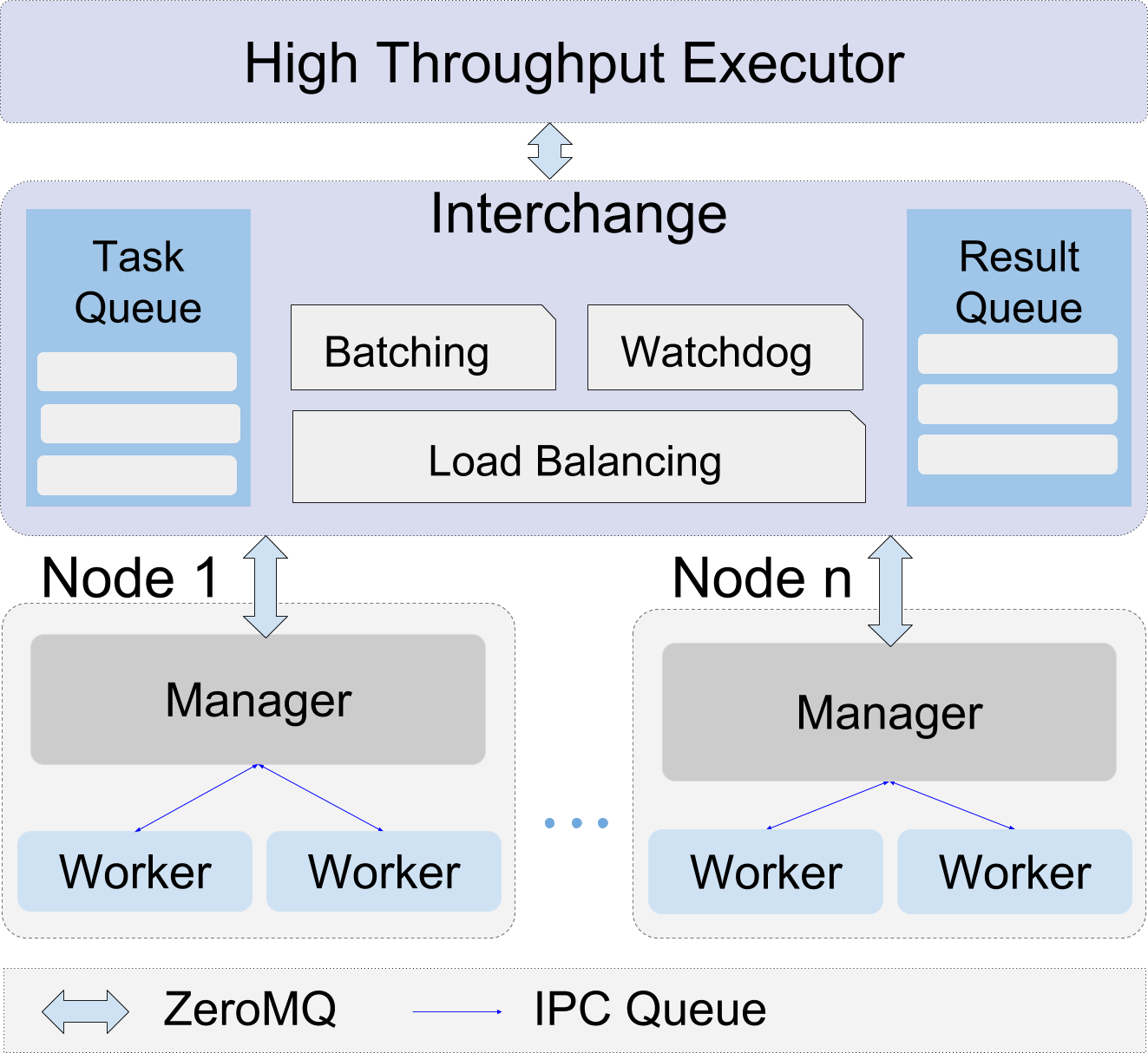}}%
\hfill
\subcaptionbox{EXEX: Extreme Scale Executor\label{fig:exex}}{\includegraphics[width=0.28\textwidth]{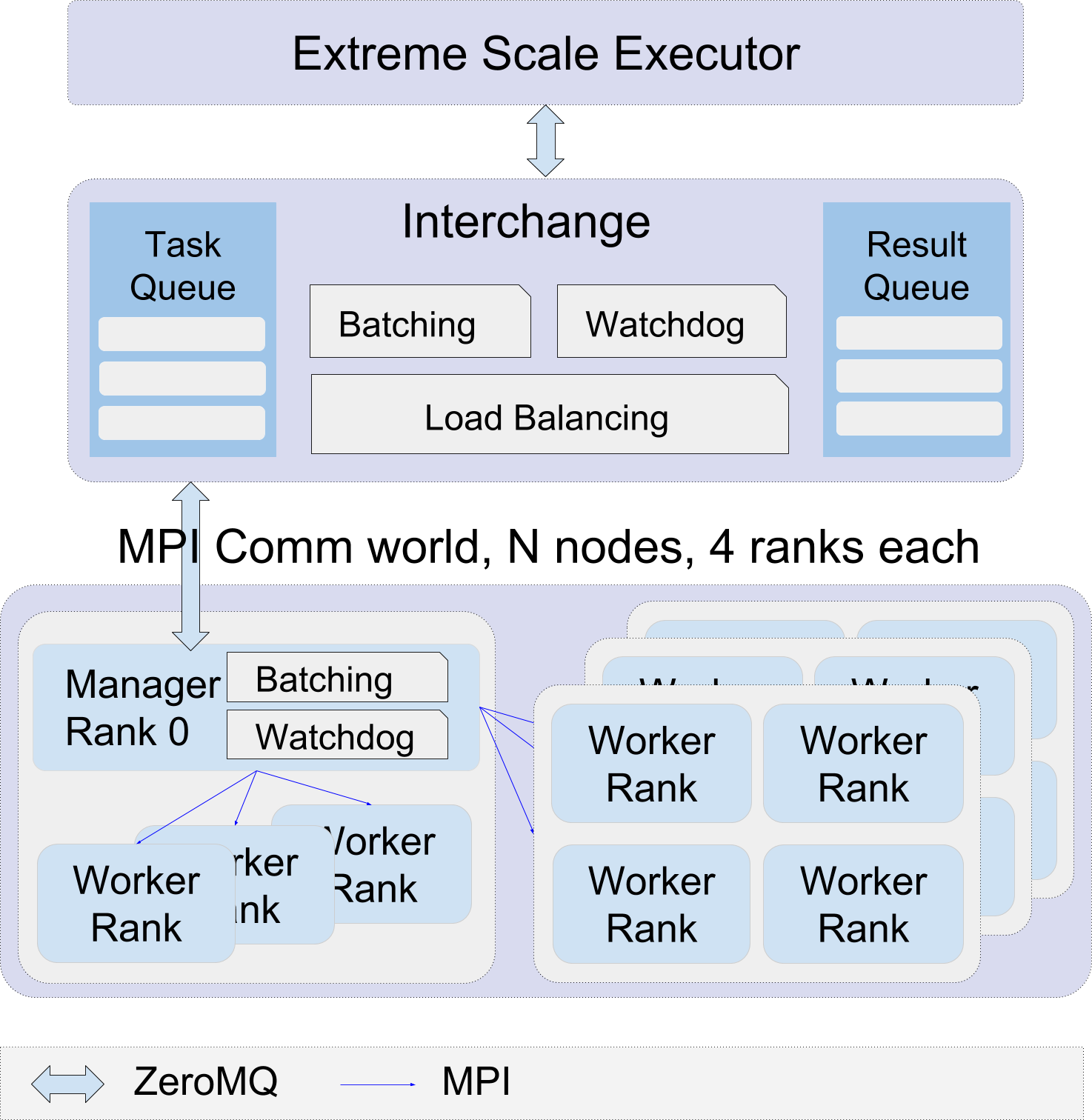}}%
\hfill
\subcaptionbox{LLEX: Low Latency Executor\label{fig:llex}}{\includegraphics[width=0.28\textwidth]{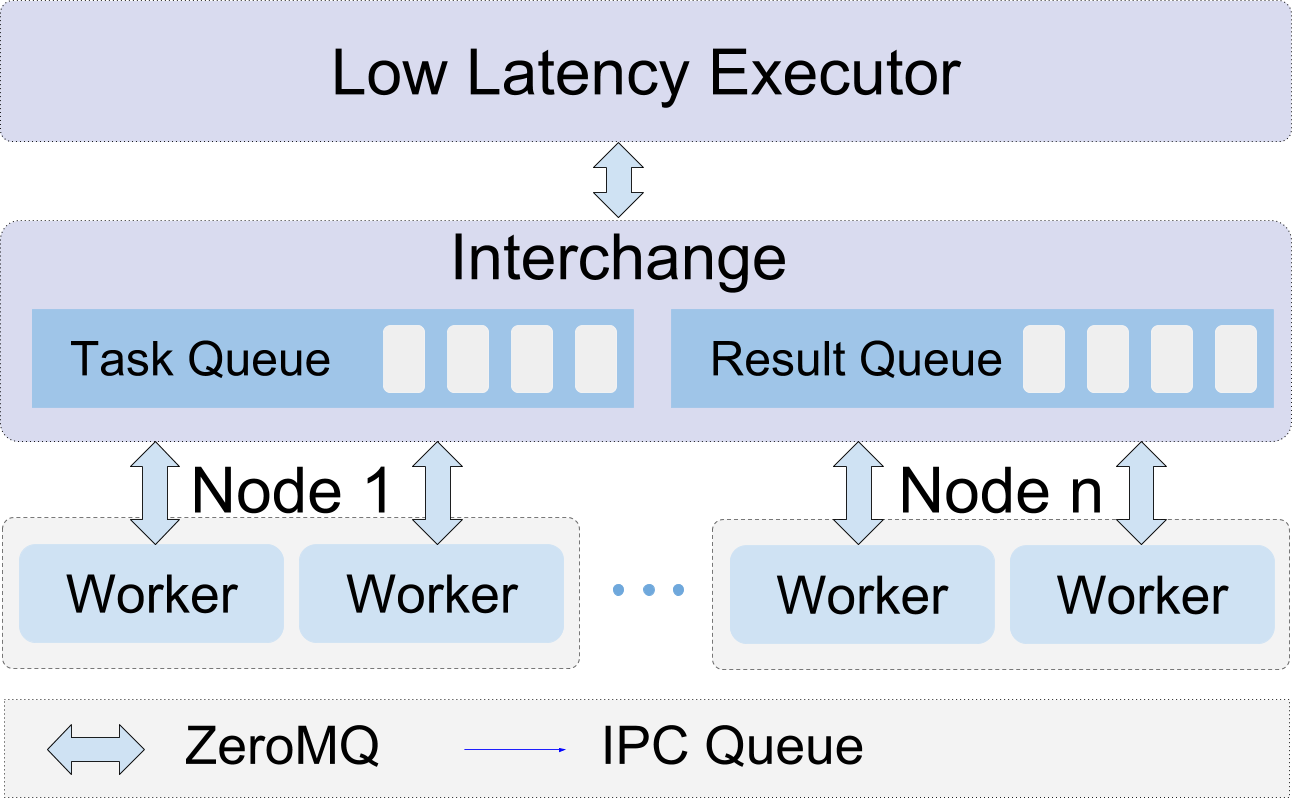}}%

\vspace{-2ex}

\caption{Architecture of Parsl's high throughput, extreme scale, and low latency executors.
\vspace{-2ex}
\label{fig:executors}}
\end{figure*}

As illustrated in \S\ref{sec:usecases}, Parsl's use cases vary widely in terms of their 
execution requirements. Individual \texttt{Apps} may run for milliseconds or days, 
and available parallelism 
can vary between none for sequential programs to millions for ``pleasingly parallel'' programs.
Executors, as the name suggests, execute Apps on one or more target execution resources such as multi-core workstations, clouds, or supercomputers.
As it appears infeasible to implement a single execution strategy that will meet so many diverse requirements
on such varied platforms,
Parsl provides a modular executor interface and a collection of executors that 
are tuned for common execution patterns. Figure~\ref{fig:executors} shows three such executors:
high throughput, extreme scale, and low latency.

Parsl executors extend the Executor class offered by Python's
\texttt{concurrent.futures} library, which allows us to use several existing solutions
in the Python Standard Library (e.g., ThreadPoolExecutor) and from other
packages such as IPyParallel~\cite{ipyparallel}.
Parsl extends the concurrent.futures executor interface
to support additional capabilities such as automatic scaling of execution
resources, monitoring, deferred initialization, and methods to set working directories.

All executors share a common execution kernel that is responsible
for deserializing the task (i.e., the App and its input arguments) and
executing the task in a sandboxed Python environment.

\subsubsection{High Throughput Executor} \label{subsection:htex}

The High Throughput Executor (HTEX) is a general-purpose executor, designed
to enable high throughput execution of tasks using a pilot job model.
It is engineered to support up to \num{2000} nodes, millions of sub-second tasks,
and multi-day workflow campaigns, all while providing a high level of fault-tolerance.
The HTEX architecture (see Figure~\ref{fig:htex}) has three major components:
executor client, interchange, and managers. 

HTEX managers (pilot agents)
are deployed onto one or more nodes by the provider. Each manager is a multi-threaded agent responsible for a
single node, initializing workers based on HTEX configuration (i.e., \texttt{workers\_per\_node}).
It advertises available capacity, and receives
batches of tasks from the interchange which are distributed to worker processes. Similarly, results are
aggregated from workers and sent to the interchange in batches. The manager
uses configurable \emph{batching} and \emph{prefetching} of tasks to minimize
communication overheads.

The interchange is a hub to which the executor client and registered managers connect 
using ZeroMQ~\cite{hintjens2013zeromq} queues.
The interchange acts as a broker, matching available tasks to managers with advertised capacity,
while using a randomized selection method to ensure task distribution fairness.

Managers and the interchange exchange periodic heartbeat messages for fault tolerance purposes.
If either party does not receive a message before a configurable threshold, the counterpart
is assumed to be lost. Managers, upon losing contact with the interchange, exit immediately
to avoid resource wastage. If the interchange detects the loss of a manager that had outstanding
tasks, an exception is sent to the executor so that DFK can make appropriate
decisions, such as scaling resources to match lost capacity and retrying failed tasks. 
HTEX also provides a separate command channel that can be used to perform administrative
actions in a synchronous fashion. For example, the interchange can be asked for outstanding
task information, to blacklist managers, or to shutdown the executor.

\subsubsection{Extreme Scale Executor}
The Extreme Scale Executor (EXEX) is designed to support the largest supercomputers---machines
with thousands of nodes, hundreds of thousands of cores, and with specialized network architectures optimized for MPI.
EXEX leverages MPI communication (using mpi4py~\cite{mpi4py}) to exploit the
highly optimized network infrastructure to manage distributed execution.

The EXEX architecture (see Figure~\ref{fig:exex}) has three major components:
EXEX executor client, interchange, and workers. EXEX is deployed as a multi-node batch job,
that uses MPI for manager-worker communication and ZeroMQ for manager-interchange communication.
EXEX uses a hierarchical task distribution model, where the managers
communicate with the interchange on behalf of workers. Upon deployment, rank 0 of the MPI
communicator takes the role of the \emph{manager}, while all other ranks assume the role of workers.

Production runs over thousands of nodes are expensive, and unfortunately the likelihood of
machine faults increase with scale. The primary drawback of using MPI as the communication
fabric is that it reduces fault tolerance in the context of many-task
applications~\cite{dorier2017supporting}. Job and node failures can result in the loss of
the entire MPI application. To alleviate these risks, we recommend that users
break their allocation into several smaller MPI worker pools within a single scheduler job.
EXEX is able to detect failures via the same heartbeat system described in \S\ref{subsection:htex}.

\subsubsection{Low Latency Executor}

The Low Latency Executor (LLEX) is designed for use cases that require low latency function execution,
but do not necessarily need high-throughput or fault-tolerance. Since the goal of LLEX is to minimize
the round-trip-time for tasks, the execution model is designed to be as minimal as possible, thus
sacrificing features such as reliability and automated resource provisioning for lower latency.

The LLEX architecture (see Figure~\ref{fig:llex}) has three major components:
executor client, interchange, and workers. To execute tasks, the LLEX client forwards task information
to the interchange, which in turn buffers and routes tasks to available workers. Results from workers
are aggregated by the interchange and returned to the client. All network communication and routing of
messages are handled by ZeroMQ. 
The interchange does not do any task tracking, and simply acts as
a relay between clients and workers. This means that the routing logic is completely stateless and
opaque to the interchange. As a result, while latency is reduced by avoiding task-tracking overhead
on the interchange, failures such as worker loss cannot be detected by LLEX.

Unlike other executors, workers connect to the interchange directly. While this design requires
a socket for each worker, message hops are reduced by one each way, reducing latency. Since tasks are short
duration, reliable execution can be guaranteed with minimal cost, even on unreliable nodes, by
\emph{timed-retries} and \emph{replication}

Finally, to meet high availability and latency requirements, LLEX assumes that it is operating on a
fixed set of compute resources. Provisioning and relinquishing resources can take
seconds to minutes on clouds and clusters, severely affecting task latencies.

\vspace{-0.05in}
\subsection{Elasticity}\label{sec:elasticity}

Workload resource requirements often vary over time. 
For example, in the map-reduce paradigm the map phase
may require more resources than the reduce phase.  
In general,
reserving sufficient resources for the widest parallelism
will result in underutilization during periods of lower load; conversely, 
reserving minimal resources for the thinnest parallelism will 
lead to optimal utilization but also extended execution time. 
Even simple bag-of-task applications may have tasks of different durations,
leading to trailing tasks with a thin workload~\cite{armstrong2010scheduling}.

Parsl implements a cloud-like elasticity model in which resource blocks
are provisioned/deprovisioned in response to workload pressure. 
Parsl provides an extensible \emph{strategy} interface by which users
can implement their own elasticity logic. By default, the elasticity
strategy can be configured with a parallelism parameter that describes
how aggressively the resources should grow and shrink in response
to waiting tasks. Given the general nature of the implementation, Parsl can
provide elastic execution on clouds, clusters, and supercomputers. 
Of course, in an HPC setting, elasticity may be complicated by queue delays.

\subsection{Data management} \label{sec:data}

Many use cases in \S\ref{sec:usecases} include Apps that
pass files to/from one another. Hard coding file paths (either local or remote)
breaks the execution location independence of a Parsl program. 
Parsl provides a \emph{file} abstraction to allow file references between
Apps. Parsl's data manager is responsible for transferring the file to
where it is needed and for transparently translating the physical location 
as needed.  

Parsl files can be defined either locally or using one of
three data access protocols: HTTP, FTP, and Globus~\cite{chard2014efficient}.
When a remote file is passed to/from an App, the Parsl
data manager first inspects the file to see if it is
available on the compute resource. 
If the file is not yet available, Parsl created a dynamic
data dependency between the App(s) that require the file as input
and a new (transparent) data transfer task. 
When the transfer is complete, the dependent App(s) are then 
able to execute. Parsl translates the file reference 
to a local path via which the App can access the file. 

The data manager performs slightly different actions depending 
on the access protocol. 
In the case of HTTP and FTP files, the data manager creates a
transfer task that is executed by the executor. That is,
it is itself a task that is executed the same way as any other task. 
Globus does not require the task to be executed on the execution
resource as it supports third party data transfer. 
When a Globus file is used, Parsl will introduce
a transfer task into the graph; however, in this case the task is executed
directly by the data manager which allows the deferment of resource provisioning until the data
has been staged to the target resource.

\subsection{Additional features}

Parsl provides a range of other features that are desirable when 
developing parallel programs. 

\textbf{Authentication}: Parsl integrates with Globus Auth~\cite{GlobusAuth}
as a ``native app.'' This allows users to authenticate in a program, 
either using interactive web-based login or cached access tokens. 
After authentication, access tokens are stored by Parsl, and these 
tokens are then used to securely access Globus Auth-enabled services (e.g., 
to transfer data or SSH to a compute resource).

\textbf{Containers}:
Parsl allows workers to be launched inside a predefined container, allowing tasks to be
executed in a customized environment. Parsl also allows containers to be used to 
execute tasks such that each invocation of a task will run a new container.
Containers provide a popular way to package and distribute software and 
to deploy it in heterogeneous environments. 

\textbf{Monitoring}: To enable both real-time and post-completion analysis and introspection of
execution information, DFK logs execution metadata and task state transitions,
and workers log task execution information, including
resource usage.  A modular DFK interface allows
monitoring information to be stored in a SQL database, Elastic Search, or files.
Logged data can be viewed via Parsl's web-based visualization interface.

\textbf{Memoization}: Parsl Apps can specify different levels of memoization
to avoid repeated execution of the same App 
with the same input arguments. Memoization can be defined at both the 
program and individual App levels. This flexibility
allows developers to select which Apps should be memoized, 
as memoization is rarely useful for non-deterministic apps.
Parsl then maintains a cache of executed Apps, function body hash, 
and arguments. 
\section{Evaluation}\label{sec:evaluation}

We evaluated the performance of Parsl with respect to latency, weak scaling, strong scaling, and elasticity using
experiments conducted on two testbeds: Midway and Blue Waters.
	
We use the ``broadwl" partition of the \textbf{Midway} campus cluster at the University of Chicago~\cite{midway},
each node of which has 28 Intel E5-2680v4 cores
running at 2.4 GHz with 64 GB RAM, interconnected with Infiniband.
The average network round trip time between two nodes was measured as 0.07 ms.

The \textbf{Blue Waters} supercomputer at the National Center for Supercomputing Applications~\cite{Bode2013} is a 13 petaFLOP Cray XE/XK hybrid system comprising \num{22636} XE compute nodes (\num{362240} cores) and \num{4288} XK compute nodes (\num{33792} cores) with an additional \num{4228} Kepler Accelerators.
We used the XE component. Each XE node has 16 AMD Interlargos cores (32 integer scheduling units) running at 2.3 GHz with 64 GB RAM, interconnected with the low-latency 3D Torus architecture. The average network round trip time between two nodes was measured as 0.04 ms.

We compare Parsl against three popular parallel computing libraries for Python.
IPyParallel (IPP)~\cite{ipyparallel} enables IPython to support parallel computing.
We compare against an IPP-based executor for Parsl (since deprecated).
FireWorks~\cite{jain2015fireworks}, a Python-based workflow management system, uses a centralized MongoDB-based LaunchPad to store tasks, and allows connected FireWorkers to query tasks from LaunchPad for execution.
Dask distributed~\cite{daskdistributed}, a framework for parallel computing in Python, has a centralized scheduler that enables task submission, makes scheduling decisions, and executes
tasks on connected workers.

\subsection{Latency}

\begin{figure}[h]
 \centering
 \includegraphics[width=0.95\columnwidth]{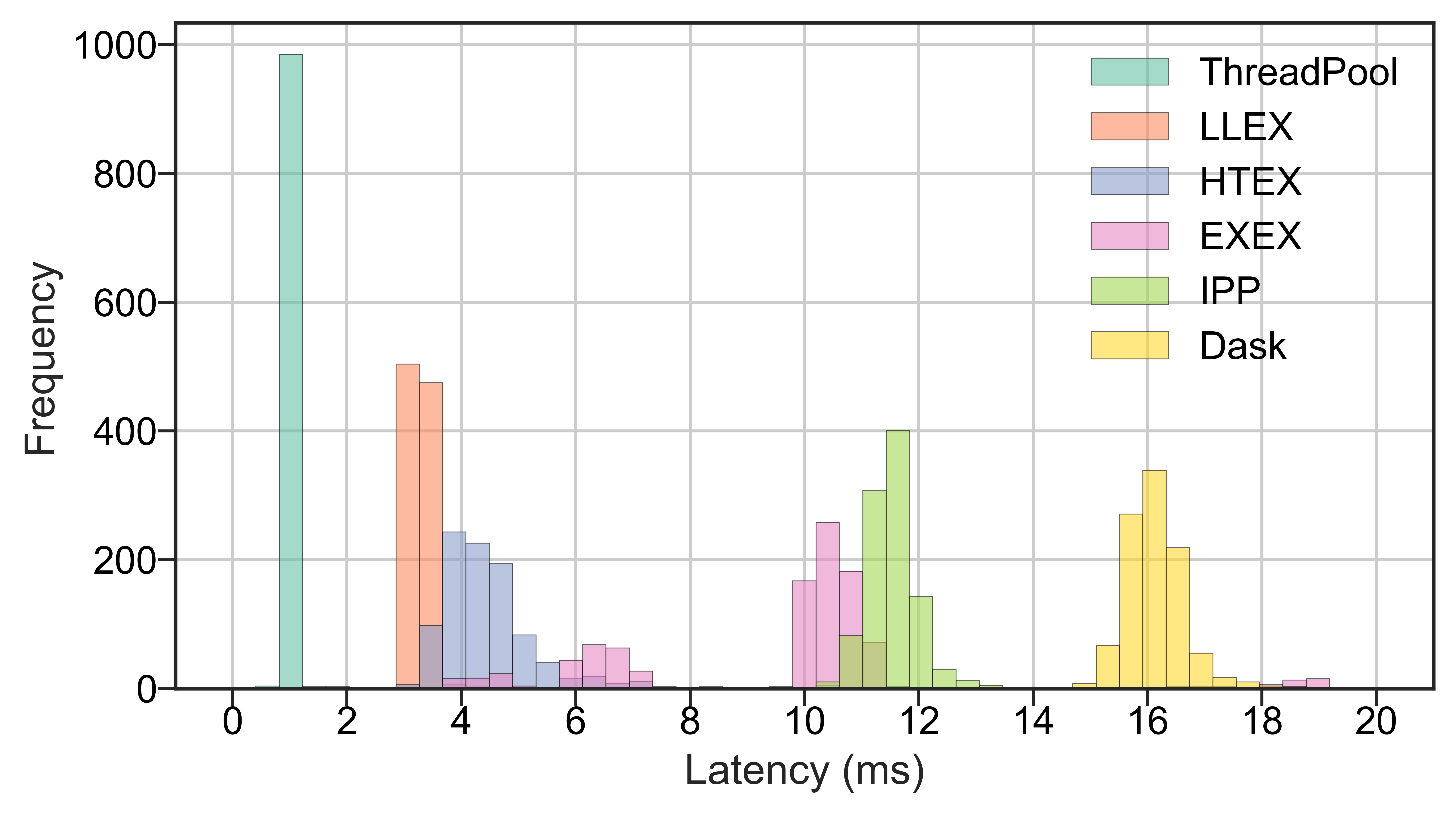}
 
 \vspace{-3ex}
 
 \caption{Distributions of task latencies when running \num{1000} tasks on Midway with different executors.}
 \label{fig:latency}
 \vspace{-2ex}
\end{figure}

We evaluated single task latency using the executors described in \S\ref{sec:executors}.
The experiment was performed on two Midway nodes:
one to run the Parsl program and the other to run a worker.
To avoid including overhead, we first deployed the worker and waited
for it to connect to Parsl.
We then launched \num{1000} tasks sequentially, recording for each the time
from submission until completion.
For comparison, we also measured execution times for the same
scenario on a single node using the ThreadPool executor,
for IPP with Parsl on two nodes,
and for Dask on two nodes. 
Figure~\ref{fig:latency} shows the distribution of task latencies in each case.

Our results show that LLEX (avg: 3.47 ms) is considerably faster
and has lower latency variability than the other executors.
LLEX is only approximately 2.43 ms slower than the local ThreadPool executor.
As expected, HTEX (avg: 6.87 ms) and EXEX (avg: 9.83 ms) exhibit larger
latencies due to the additional complexity of their respective executor architectures.
The Parsl executors all have lower latencies than IPP (avg: 11.72 ms) and Dask (avg: 16.19 ms).

\begin{figure*}
	\centering
	\includegraphics[width=1\textwidth]{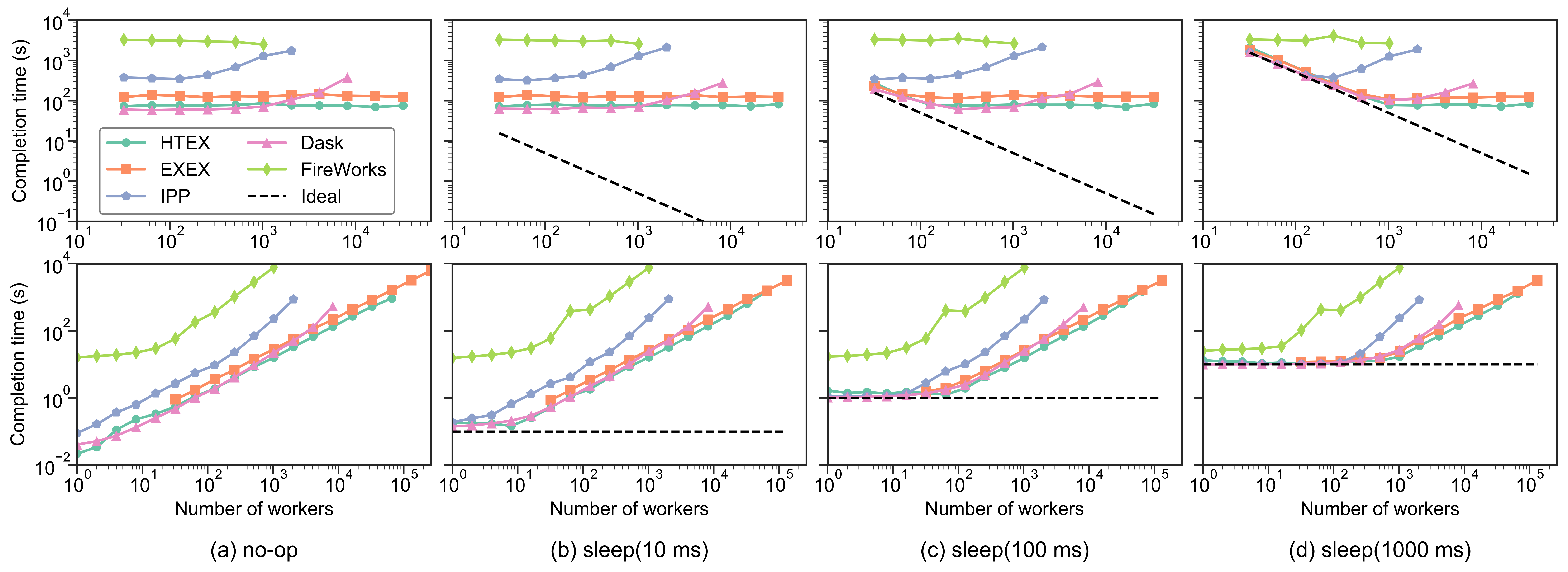}
	\vspace{-5ex}
	\caption{Top row: time to execute \num{50000} tasks over all workers (strong scaling). Note: FireWorks results were obtained using \num{5000} tasks over all workers. Bottom row: time to execute 10 tasks per worker (weak scaling). For each row, plots are for (from left to right) tasks of 0, 10, 100, and 1000 ms. Legend is at top left.}\label{fig:scaling}
\end{figure*}

\subsection{Scalability}
We studied strong and weak scaling on Blue Waters.
In strong scaling, the total problem size is fixed; in weak scaling, the problem size \emph{per CPU core} is fixed.
In both cases, we measure completion time as a function of number of CPU cores.
An ideal framework should scale linearly,
which for strong scaling means that speedup scales with the number of cores,
and for weak scaling means that completion time remains constant as the number of cores increases.

To measure the strong and weak scaling of Parsl executors, we created Parsl programs to run tasks with different durations,
ranging from a ``no-op''---a Python function that exits immediately---to tasks that sleep
for 10, 100, and \num{1000} ms.
For each executor we deployed a worker per core on each node.

We compare the scalability of Parsl with IPP on Parsl, FireWorks, and Dask distributed.
We launched the LaunchPad/scheduler of Fireworks and Dask distributed on one compute node, and the workers on the other compute nodes. For fair comparison, we deployed each worker process on one core and disabled caching (if available).

\textbf{Strong scaling.}
We launched \num{50000} independent tasks of different durations (no-op and 10, 100, \num{1000} ms) on an increasing number of workers and with different Parsl executors as well as IPP, and Dask distributed. For FireWorks we only launched \num{5000} tasks due to the limited allocation available to us on Blue Waters. 
Notice that the measurements for ``no-op" tasks essentially reflect the overhead of the executor.

The top row of Figure~\ref{fig:scaling} show the strong scaling results.
HTEX provides the best performance in all cases, slightly
exceeding what is possible with EXEX, while EXEX scales to significantly
more workers than the other executors and frameworks.
Encouragingly, both HTEX and EXEX remain nearly constant, indicating that they
likely 
will continue to perform well at larger scales.
In comparison with the other frameworks, 
FireWorks has the highest overhead even with only \num{5000} tasks: almost an
order of magnitude greater than the other executors/frameworks.
Both IPP and Dask distributed exhibit a similar trend of increasing
overhead as the number of workers increases beyond 512.
Dask distributed slightly outperforms HTEX and EXEX when there are
fewer than 1024 workers.

\textbf{Weak scaling.}
Here, we launched 10 tasks per worker,
while increasing the number of workers.
(We limited experiments to 10 tasks per worker, as on \num{3125} nodes,
that already represents \num{3125} nodes $\times$ 32 workers/node $\times$ 10 tasks/worker, or 1M tasks.)
The bottom row of Figure~\ref{fig:scaling} shows our results.
We observe that HTEX and EXEX outperform other executors and frameworks with more than 4096 workers (128 nodes).
While all frameworks exhibit similar trends,
with completion time remaining close to constant initially and increasing rapidly as the number of workers increases,
some executors/frameworks exhibit sublinear scaling more quickly than others.
FireWorks scales sublinearly from around 32 workers, IPP at 256 workers (8 nodes),
and Dask distributed, HTEX, and EXEX at \num{1024} workers (32 nodes).

\textbf{Maximum number of workers.}
To further investigate scalability we now consider the maximum number of workers that
can be connected without failure. To do so, we configured the executors on Blue Waters and
continued to add workers until we observed errors.
The maximum number of connected workers we observed are summarized in Table~\ref{table:maxWorkers}.
We were able to launch \num{2048} workers on 64 nodes with IPP,
\num{65536} workers on \num{2048} nodes with HTEX, and
\num{262144} workers on \num{8192} nodes with EXEX.
For HTEX and EXEX we were not able to force an error and were instead
limited by the number of nodes we could provision in our allocation.
Dask distributed scaled to \num{8192} workers on 256 nodes, after which we observed connection failures
due to the fact that each worker must connect to the centralized scheduler,
which can handle only a limited number of connections.
FireWorks scaled to \num{1024} workers on 32 nodes,
although at this point we observed slow performance and a variety of errors,
such as time outs from its 
MongoDB server.

\subsection{Throughput}
We measured the maximum throughput of all the Parsl executors,
as well as IPP with Parsl, Dask distributed, and FireWorks, on Midway.
To do so, we ran \num{50000} ``no-op" tasks on a varying number of workers and recorded the completion times.
The throughout is computed as the number of tasks divided by the completion time.

As shown in Table~\ref{table:maxWorkers},
IPP, HTEX, and EXEX achieved maximum throughputs of 330, \num{1181}, and \num{1176} tasks/s, respectively. 
Dask distributed had the highest throughput, of \num{2617} tasks/s, likely as it
is optimized for short duration jobs on small clusters. FireWorks
had the lowest throughout due to its slow centralized database.

\begin{table}[h]
	\centering
	\small
	\vspace{-0mm}
	\begin{threeparttable}
		\caption{Capabilities and capacities of different Parsl executors and other parallel Python tools.\vspace{-3ex}}
		\vspace{0mm}
		\begin{tabular}{| c| c|c|c|}
			\hline
			\multirow{2}{*}{\textbf{Framework}} & \textbf{Maximum}   & \textbf{Maximum} & \textbf{Maximum} \\
			& \textbf{\# of workers}\tnote{\dag} & \textbf{\# of nodes}\tnote{\dag} & \textbf{tasks/second}\tnote{$\ddagger$} \\ \hline
			Parsl-IPP & \num{2048} & 64  & 330 \\ \hline
			Parsl-HTEX & \num{65536} & \num{2048}\tnote{*}  & \num{1181}\\ \hline
			Parsl-EXEX & \num{262144} & \num{8192}\tnote{*}  & \num{1176}\\ \hline
			FireWorks & \num{1024} & 32   & 4\\ \hline
			Dask distributed & \num{8192} & 256  & \num{2617}\\ \hline
		\end{tabular}
		\begin{tablenotes}\small
			\item[*] Limited by the the number of nodes we could allocate on Blue Waters during our experiments; this is not a scalability limit.
			\item[\dag] These results are specific to Blue Waters, one core per worker, and using default configuration as in each framework's documentation.
			\item[$\ddagger$] The results in this column are collected on Midway.
		\end{tablenotes}
		\label{table:maxWorkers}\vspace{-0in}
	\end{threeparttable}
\end{table}

\subsection{Elasticity}

\begin{figure}[h]
 \centering
 \includegraphics[width=0.95\columnwidth,trim=0 0in 0 0in, clip]{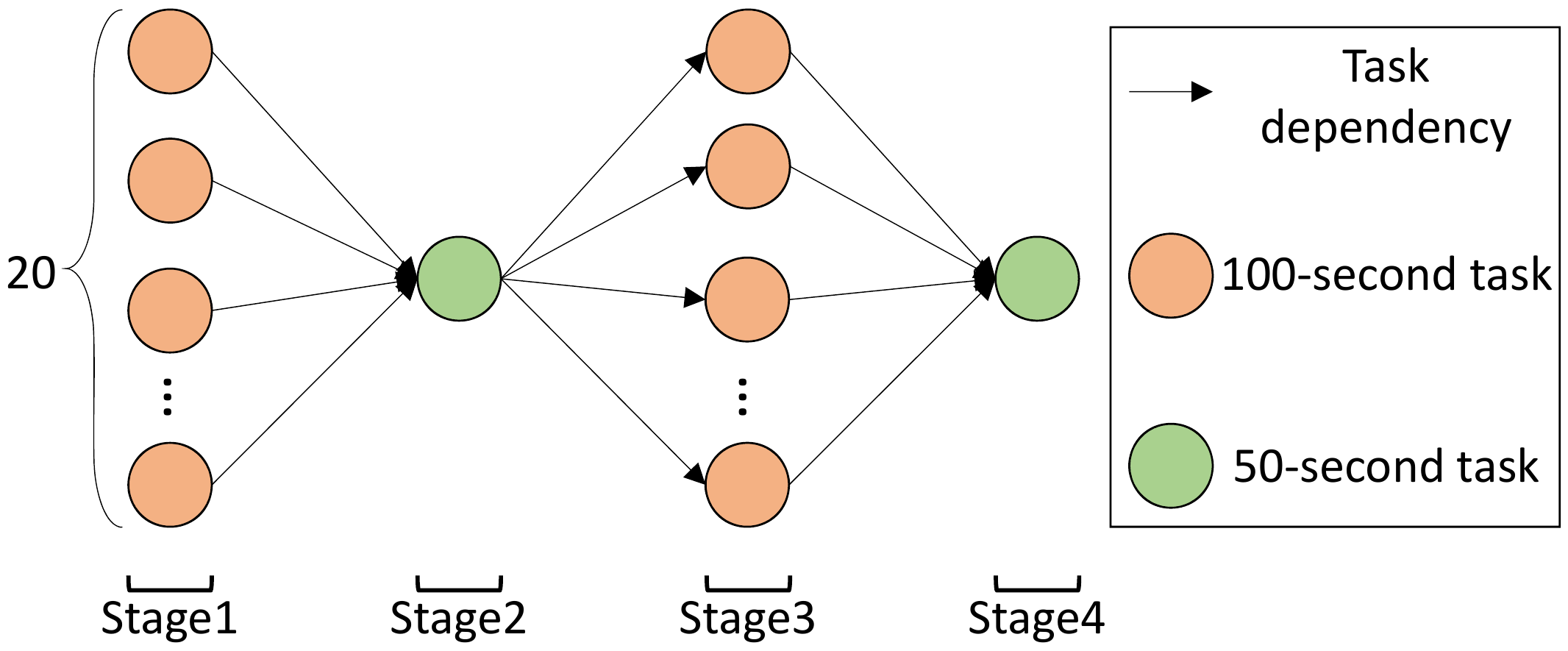}

 \vspace{-2ex}

 \caption{Workflow graph used in elasticity study.}
 \label{fig:elasticity-workflow}
 \vspace{-2ex}

\end{figure}

We used the four-stage workflow shown in Figure~\ref{fig:elasticity-workflow} to
study the efficacy of Parsl's elastic resource management.
The first and third stages have the widest parallelism, with 20 tasks, while the
second and fourth stage are reduce-like stages with a single task each.
Every task is a sleep task and expends the capacity of a single worker for the specified duration (100 or 50 s).
We executed this workflow on Midway with and without elasticity enabled and report the overall worker utilization of the acquired
resources, calculated as the ratio of total wall clock time of tasks to that of the workers.

\begin{figure}[h]
 \centering
 \includegraphics[width=0.95\columnwidth]{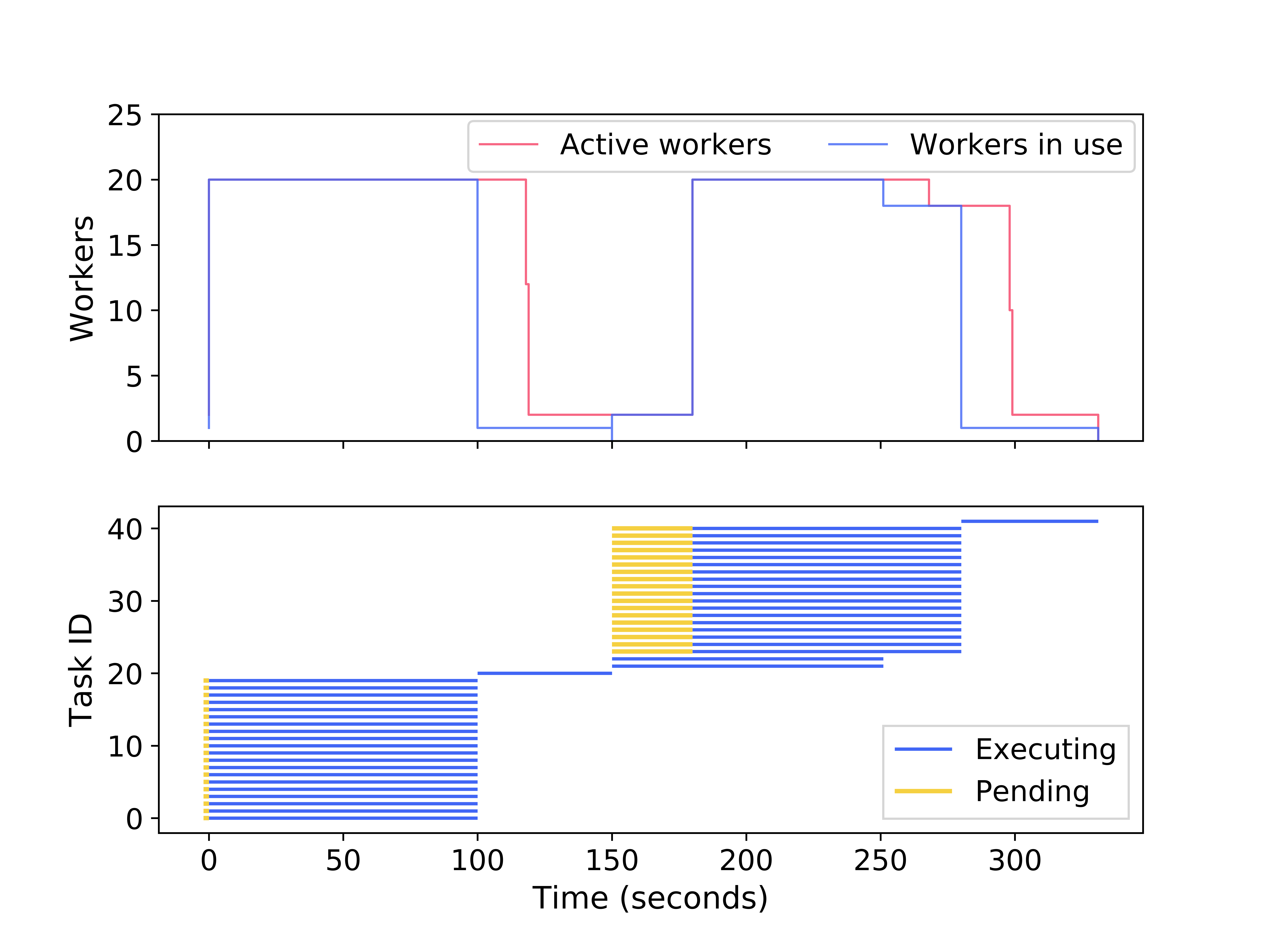}
 
 \vspace{-2ex}
 
 \caption{Utilization with elasticity. Above: Number of workers 
active and in use over time. Below: Task lifecycle, including both time
waiting in a queue and time executing.}
 \label{fig:cpu-utilization-elasticity}
  \vspace{-2ex}
\end{figure}

Figure~\ref{fig:cpu-utilization-elasticity} shows worker utilization when elasticity is enabled.
Without elasticity, we observe average worker utilization of 68.15\% and a makespan of 301 s. 
The worker utilization is poor during the reduce stages, wasting computing resources.
In contrast, with elasticity enabled, average worker utilization is 84.28\%,
because Parsl can scale the number of blocks used dynamically, based on the workload.
The makespan is slightly increased to 331 s.  Overall, in this example, utilization
is increased by 23.6\% at the expense of a 9.9\% increase in makespan.

\section{Related work}\label{sec:related}

\textbf{Analytics platforms.}
Hadoop~\cite{hadoop} and Spark~\cite{spark} are popular data-parallel systems for data analytics. 
Both follow the map-reduce model and are primarily designed for I/O-intensive applications, 
such as sorting, counting and aggregation.
In contrast, Parsl implements a more general approach to parallelism, enabling various types
of parallelism to be expressed in Python.

\textbf{Scientific workflow engines.}
Many workflow systems---238 at the 
time of writing~\cite{workflowsystems}---enable the orchestrated execution of multiple applications. 
Examples are
Pegasus~\cite{pegasus}, Galaxy~\cite{galaxy}, Swift~\cite{swift}, NextFlow~\cite{di2017nextflow}, FireWorks~\cite{jain2015fireworks}, Apache Airflow~\cite{airflow}, and Luigi~\cite{luigi}.

Pegasus and Galaxy implement a static DAG model in which users define a 
DAG, using an XML document or GUI, and subsequently execute that DAG. 
The Common Workflow Language (CWL)~\cite{cwl} attempts to standardize workflow
descriptions using a YAML specification. 
These workflow systems take a different approach to addressing
parallelism, requiring static, upfront definition of a workflow. Parsl 
instead augments Python, offering the full power of the Python programming
language to create dynamic, parallel applications. 

Swift and NextFlow rely on custom DSLs to express parallelism. While
they provide for excellent performance; they have a steep learning 
curve, and a limited set of programming constructs from which to create
programs.  

Python-based workflow systems such as FireWorks, Airflow, and Luigi
enable the explicit description of dependency graphs in Python, rather than the ability
to augment Python with parallelism. FireWorks focuses on fault-tolerance, rather
than scalability and performance, relying on a persistent MongoDB to communicate task state.
Airflow relies on a centralized scheduler that processes the task graph and manages
execution on connected workers. Luigi, requires parallelism to be represented
in classes, where a task describes its explicit input/output objects. When executed, 
Luigi builds a graph by introspecting the connected classes.

\textbf{Parallel computing in Python.}
PyDFlow~\cite{PyDFlow} was one of the first Python libraries to offer lazy evaluation in 
Python.  Like Parsl, PyDFlow allows developers to wrap Python functions with decorators to denote lazy evaluation semantics. These functions produce I-vars~\cite{id} which, when accessed, produce
a task graph that can be evaluated concurrently. Our experiences with PyDFlow have informed
the development of Parsl.

The Dask~\cite{dask} Python library supports parallel computing for 
data analytics. Unlike Parsl, Dask focuses on implementing parallel
versions of common Python libraries, such as NumPy, Pandas, and Scikit-learn.
Dask also offers low-level APIs for composing custom parallel systems,
with constructs such as ``delayed'' for wrapping function calls, and
futures for developing asynchronous programs. 
Dask distributed~\cite{daskdistributed} extends Dask's execution model to support distributed execution
on small clusters. It relies on a centralized scheduler that coordinates task 
submission and dynamic scheduling across multiple nodes.
While Parsl and Dask share common features, Dask is primarily focused
on data parallelism via high level libraries and on local or small-scale distributed
execution environments.

Ray~\cite{moritz2018ray} is a distributed system designed to support 
training, serving, and simulation for reinforcement learning applications.
Ray can execute millions of short-duration tasks. 
To achieve this performance it relies on a distributed scheduler
and a distributed Redis-based fault-tolerant metadata store.
The global control store maintains the entire state of the system, allowing
the distributed scheduler to make rapid scheduling decisions based on global state. 
Ray implements a unified interface that enables expression of both
actor and task-parallel abstractions for representing parallelism. 

PyCOMPSs~\cite{tejedor2017pycompss},
a Python interface around the 
COMPSs system,
is a framework for parallel computing in Python. 
As with Parsl, users decorate functions with constructs to aid workflow assembly 
and execution.
COMPSs is responsible for interpreting the task graph and scheduling tasks to available resources. 

While these frameworks implement a task graph model, Parsl focuses on a broader
problem of enabling parallelism in Python. Parsl therefore tackles problems
that range from many short tasks through to long tasks executing at extreme scale. 
The underlying model employed by Parsl could be used by many of these comparable
frameworks to enable parallelism at a higher level. 

\textbf{Machine learning frameworks.}
TensorFlow~\cite{abadi2016tensorflow} focuses on machine learning applications and can achieve high performance for   linear algebra and other numerical computations.
It represents computation as a dataflow graph, mapping each graph node to different machines or computational units (e.g., CPU and GPU).
TensorFlow provides little support for more general parallelism, task composition, 
or other execution models. 
\section{Summary}\label{sec:summary}

Parsl addresses two major trends in programming: the increasing
use of high level languages, such as Python, to compose rather
than write software; and the growing need for parallel computing in
analysis and simulation.
Parsl allows parallelism to be expressed via the use of
simple decorators that enable safe,
deterministic parallel programs; supports scalable execution
from laptops to supercomputers; and provides a flexible architecture that
can address the varied requirements of scientific analyses.

Our performance studies highlight the unique position in the parallel Python ecosystem that Parsl fills.
Systems like FireWorks support use cases that require concurrent execution
of few ($<$\num{1000}) long-running tasks ($>$100 s)~\cite{jain2015fireworks}. 
Dask distributed can manage short tasks efficiently, but
is designed for small-scale cluster deployments of fewer than 100 nodes.
Parsl, and its flexible executor model, effectively fill the unmet needs of a variety of
use cases, enabling efficient scalability up to $\sim$\num{8000} nodes (and likely
more if allocations permit), execution overhead of less than 5 ms, and high-throughput execution
of $\sim$\num{1200} tasks per second.
We provide guidelines for selecting
Parsl executors in Figure~\ref{fig:executor_selection}.

Our future work focuses on expanding Parsl capabilities. 
Having developed a flexible and general-purpose parallelism library,
we next aim to investigate constructs for delivering parallelism such
as maps and additional synchronization primitives such as barriers. 
We are particularly interested in supporting parallelism 
in higher-level libraries and domain-science libraries. 
We are also working to expand Parsl's data
management capabilities, including by enabling direct data staging between
nodes, ephemeral caching of data on nodes, and optional sandboxing 
environments. 

Parsl is 
an open source project
available on GitHub: \url{https://github.com/Parsl/parsl}. Community contributions are welcome.

\begin{figure}
\begin{fminipage}{\linewidth}
\small
\begin{description}[leftmargin=8mm] 
\item[\textbf{LLEX}]
for \textbf{interactive} computations on \textbf{$\le$10} nodes.
\item[\textbf{HTEX}]
for \textbf{batch} computations on \textbf{$\le$\num{1000}} nodes. (\emph{For good performance, task-duration / \# nodes  $\ge$ 0.01: e.g., on 10 nodes, tasks $\ge$ 0.1 s.})
\item[\textbf{EXEX}]
for \textbf{batch} computations on \textbf{>\num{1000}} nodes.
(\emph{For good performance, task durations $\ge$ 1 min.})
\end{description}
\end{fminipage}

\vspace{-2ex}

\caption{Guidelines for selecting Parsl executors.
  \label{fig:executor_selection}}
\vspace{-3ex}
\end{figure}

\section*{Acknowledgment}
This work was supported in part by the NSF (ACI-1550588)
and DOE (DE-AC02-06CH11357).
It relied on the Blue Waters sustained-petascale computing project, which is supported by the NSF (OCI-0725070, ACI-1238993) and the State of Illinois.

\bibliographystyle{ACM-Reference-Format}
\bibliography{refs}

\end{document}